\documentclass[12pt,a4paper]{article}
\usepackage{anysize}
\usepackage{amsmath}
\usepackage{graphicx}
\title{A note on the ABC-PRC algorithm of Sissons \emph{et
al.}(2007)}
\author{Mark A. Beaumont}
\begin{document}
\maketitle

\section{The ABC-PRC Algorithm}

\begin{figure}[tb]
\centering
\includegraphics[width=7in]{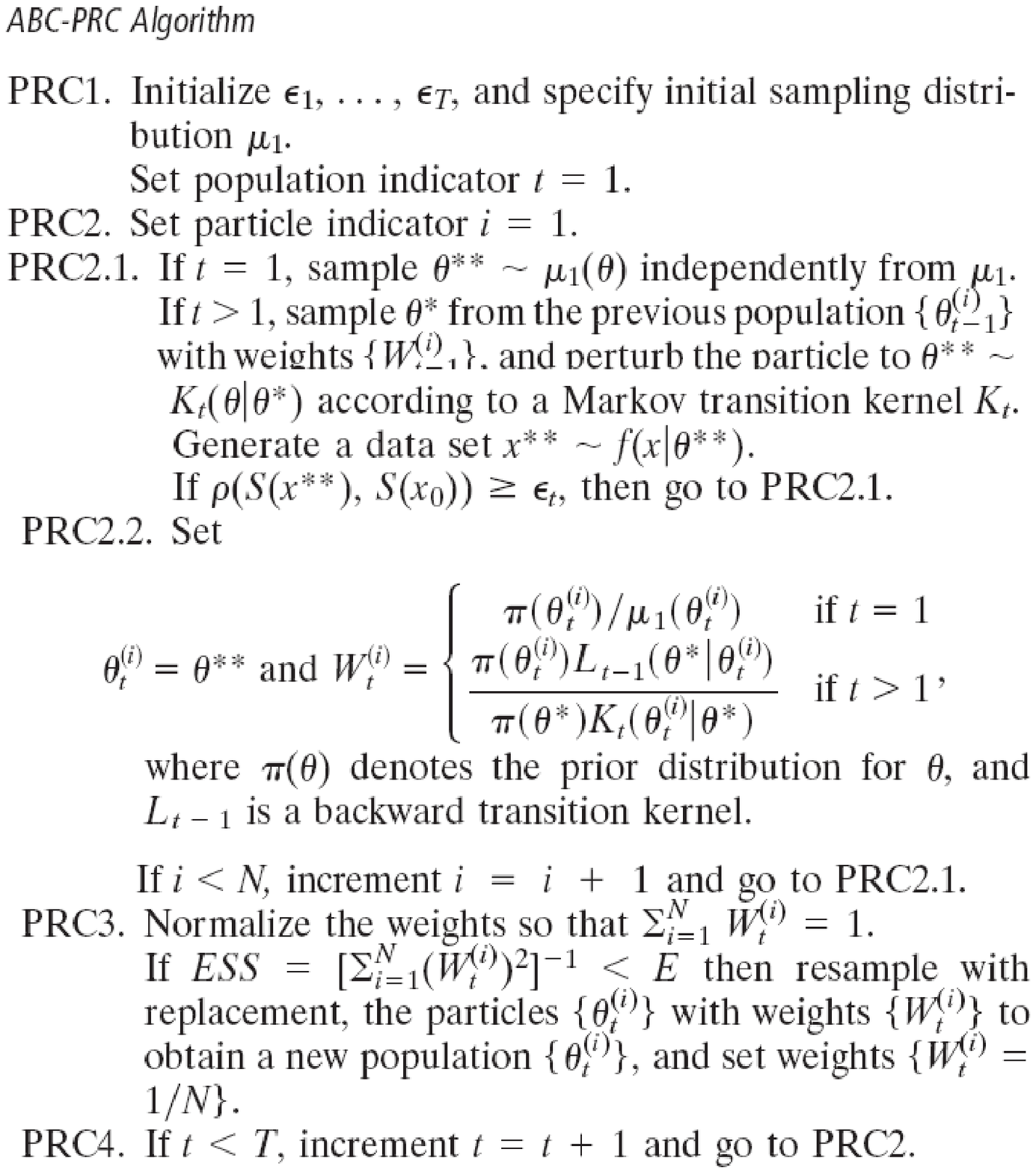}
\caption[]{The ABC-PRC algorithm of Sissons \emph{et al.}, PNAS,
2007}\label{fig:sissons-alg}
\end{figure}

A sequential Monte Carlo method for performing approximate Bayesian
Computation (``Monte Carlo without likelihoods'') has been proposed
by Sissons, Fan and Tanaka (PNAS, 2007). The main algorithm that is
used in their paper is given here in Figure \ref{fig:sissons-alg}.
In this algorithm $N$ is the number of particles $\theta$ is a
vector of parameters, the $\epsilon_t$ are tolerances such that if
$\rho(S(x),S(x_0)) \leq \epsilon$ the simulated summary statistics,
$S(x)$ are considered `near enough' to the target summary statistics
$S(x_0)$, where $\rho()$ is some distance function. The authors
state
\begin{quote} \small Samples $\{ \theta_T^{(i)} \}$ are weighted samples
from the posterior distribution $f(\theta | \rho(S(x),X(x_0)) \leq
\epsilon)$.\end{quote} They also state
\begin{quote} \small
Finally we note that if $K_t(\theta_t | \theta_{t-1}) =
L_{t-1}(\theta_{t-1} | \theta_t) $, $\mu_1(\theta) = \pi(\theta)$
and the prior $\pi(\theta) \propto 1$ over $\Theta$, then all
weights are equal throughout the sampling process and maybe safely
ignored...\end{quote}

\section{Problematic Aspects}

Intuitively the simplified algorithm, which arises from making the
assumptions in the latter statement above, is rather puzzling
because it is unclear what corrects for the fact that one is
sampling from a distribution that, if the transition kernel
variance, $K_t(\theta_t | \theta_{t-1})$, is small, is progressively
moving away from the prior.

In the algorithm given in Figure \ref{fig:sissons-alg} there is the
statement: if $\rho(S(x**),S(x_0)) \geq \epsilon_i$, then go to
PRC2.1. Intuitively this makes the algorithm superficially appear as
$N$ parallel MCMC-ABC chains, with the exception that there is
resampling among the chains at each iteration. Certainly if one
takes the special case of $N=1$ then it looks similar to the
ABC-MCMC algorithm of Marjoram et al. (2003). However a crucial
difference is that in the ABC-MCMC the current value is the old
value if the update is not accepted, otherwise it is the new value,
whereas in this SMC algorithm you keep going until you get an
acceptance. So in the ABC-MCMC algorithm, you are guaranteed that if
the point you update is already from the posterior distribution, the
next point will also be from the posterior (as given by the proof,
based on satisfying detailed balance, in Marjoram et al, 2003, page
15325), whereas in the SMC this is not the case. To see this,
imagine that the kernel chosen has a very small variance (close to
0, in fact). In the ABC-MCMC, following the proof of detailed
balance in Marjoram et al (2003), we are guaranteed that if
$\theta_i$ is drawn from the posterior, then $\theta_{i+1}$ is also
drawn from the posterior (whatever variance of the kernel is
chosen). In the SMC, for example, imagine that one has only two
resampling steps, and imagine that a sample is taken with $\epsilon$
close to 0, and kernel variance close to 0. The first step gives you
(almost) the posterior distribution , $\theta_i \tilde p(\theta | x
= x_c)$. The next step, since the prior is now actually the
posterior, is exactly like performing standard rejection with such a
prior, and gives you (almost) $[p(\theta | x = x_0)]^2)$, the square
of the posterior distribution, and so on, progressively. Thus,
initially at least, the posterior distribution generated at the
$i$th step is an increasingly poor estimate of $[p(\theta | x)]^i)$.

\section{Examples}

As a toy example consider the case of computing the posterior
distribution of the parametric mean, $\mu$ of a Gaussian
distribution with a known variance $\sigma^2$, given a vector of $n$
observations $y$. The prior for $\mu$ is taken to be Gaussian with
mean $\mu_0$ and variance $\tau^2$. The posterior $p(\mu | y)$ is
then given as
$$
N \left ( \frac{\frac{\mu_0}{\tau^2} +
\frac{n*\bar{y}}{\sigma^2}}{\frac{1}{\tau^2} + \frac{n}{\sigma^2}},
\frac{1}{\frac{1}{\tau^2}+\frac{n}{\sigma^2}} \right ),
$$
where $\bar{y}$ is the arithmetic mean of the elements of $y$, is a
sufficient statistic for this problem, and is used as the only
statistic for the ABC analyses described here. In order to test the
simplified version of the ABC-PRC algorithm above (although the
argument applies also to the more complicated version), I let
$\tau^2 \rightarrow \infty$ so that a flat improper prior is
assumed, in which case one obtains
$$
N \left ( \bar{y}, \frac{\sigma^2}{n} \right ).
$$

\begin{figure}[tb]
\includegraphics[width=6in]{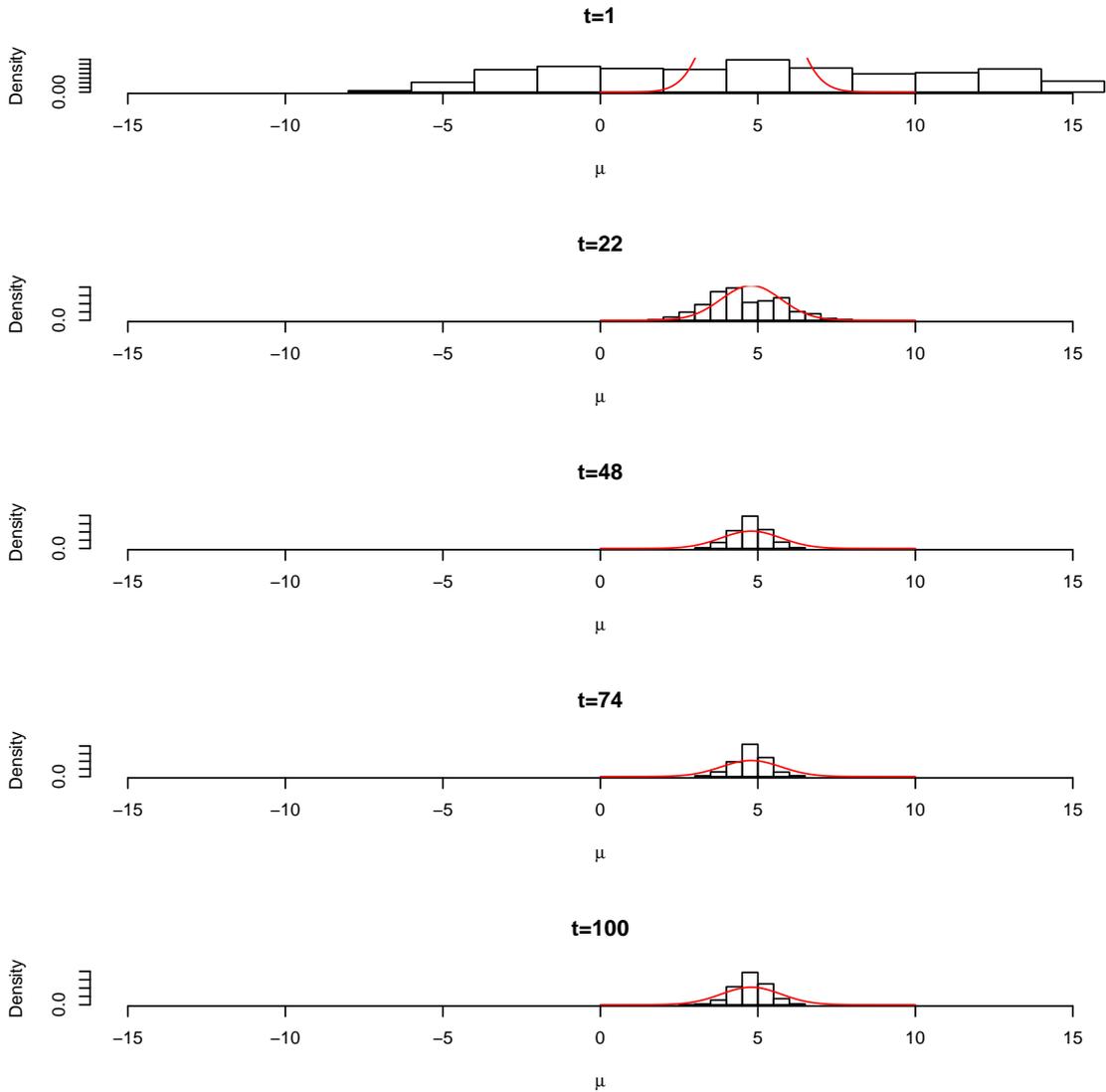}
\caption[]{Application of the ABC-PRC algorithm
 with kernel variance of 0.1}\label{fig:small-kern}
\end{figure}

Figure \ref{fig:small-kern} indicates the application of the ABC-PRC
algorithm of Sissons \emph{et al.} (2007) to data with
$\bar{y}=4.786624$ and $n = 10$. The known variance, $\sigma^2$ is
set at 9, and so the posterior variance should be 0.9, with
posterior mean of 4.786624. To approximate a flat uniform prior, a
uniform with bounds (-15,15) is used for the initial sample.
The ABC-PRC algorithm was run for 100 iterations. The schedule of
tolerances is given below:\\
\begin{center}
\begin{tabular}[t]{ll}
$t$ & $\epsilon_t$  \\ \hline
1--10 & 10.0 \\
11--20 & 5.0\\
21--30 & 2.0\\
31--40 & 1.0 \\
41--50 & 0.5 \\
51--60 & 0.2 \\
61--70 & 0.1 \\
71--80 & 0.05 \\
81--90 & 0.02 \\
91--100 & 0.01.
\end{tabular}
\end{center}

\begin{figure}[tb]
\centering
\includegraphics[width=4in]{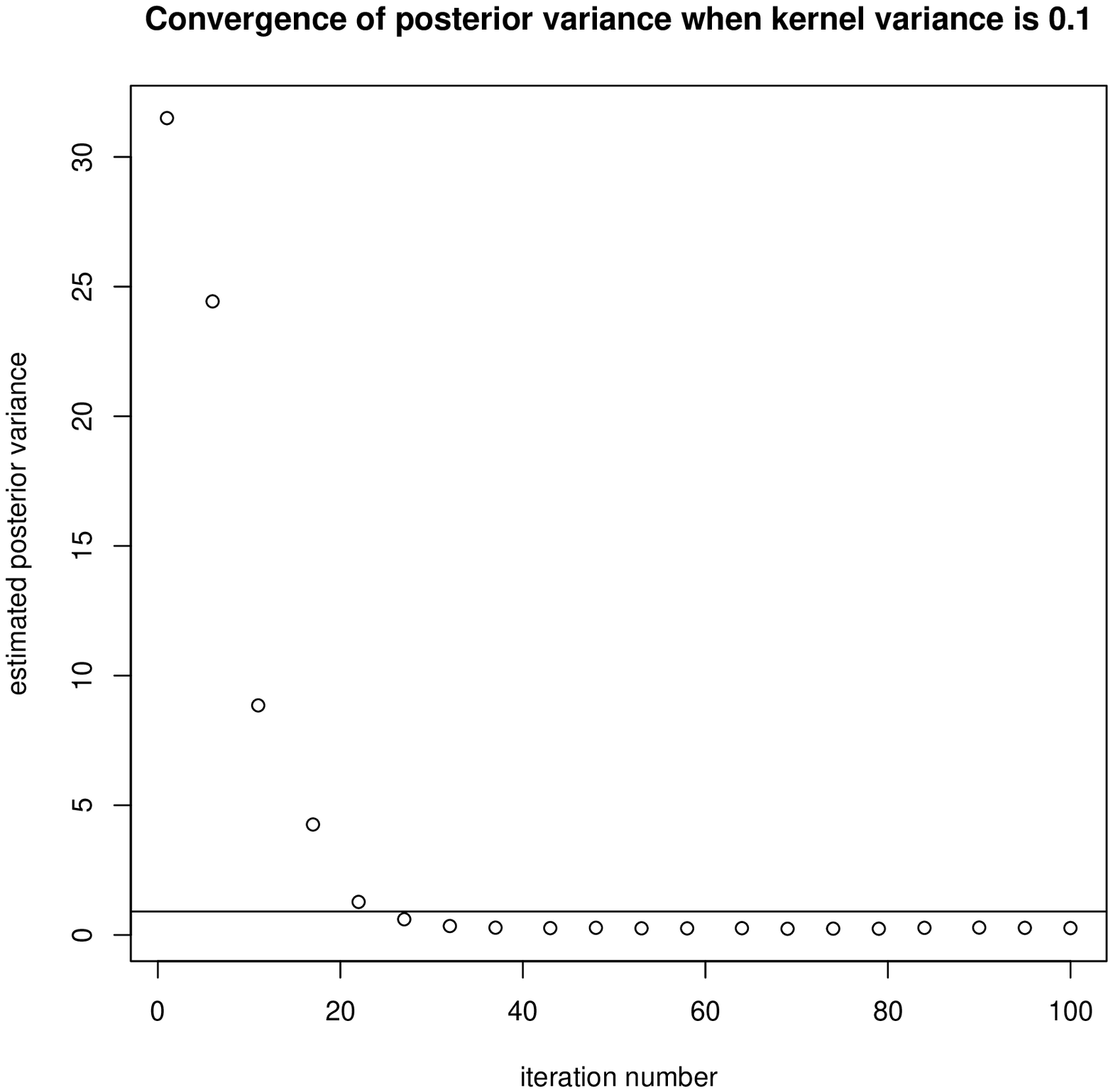}
\caption[]{Application of the ABC-PRC algorithm
 with kernel variance of 0.1. The horizontal line shows
 the theoretical value.}\label{fig:small-kern-converge}
\end{figure}

The method converges smoothly, as indicated in Figure
\ref{fig:small-kern-converge}, but to a variance that is too small.
The true posterior variance is 0.9. The posterior variance to which
the ABC-PRC method appears to converge is around 0.26.

\begin{figure}[tb]
\includegraphics[width=6in]{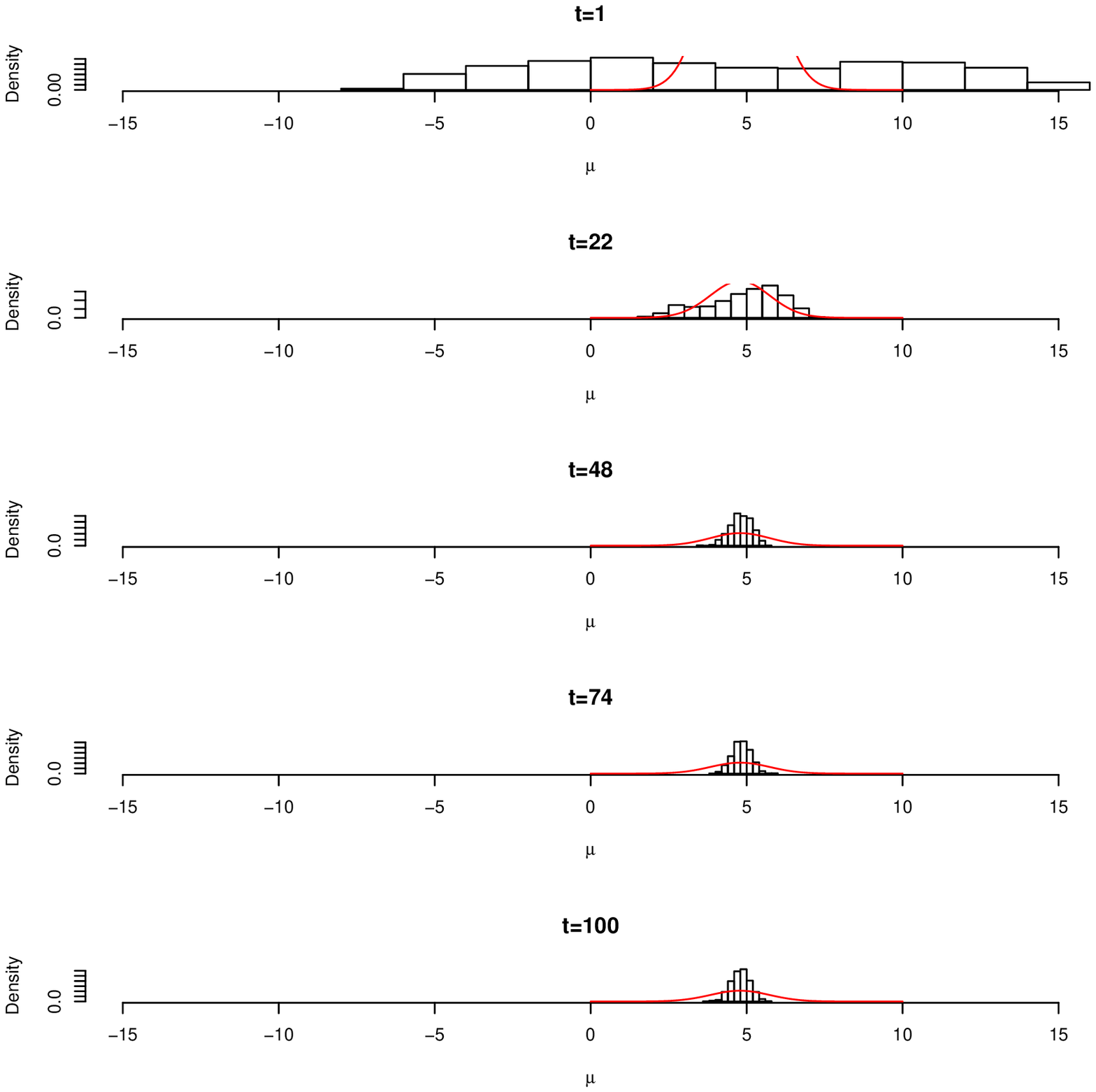}
\caption[]{Application of the ABC-PRC algorithm
 with kernel variance of 0.01}\label{fig:smaller-kern}
\end{figure}

\begin{figure}[tb]
\centering
\includegraphics[width=4in]{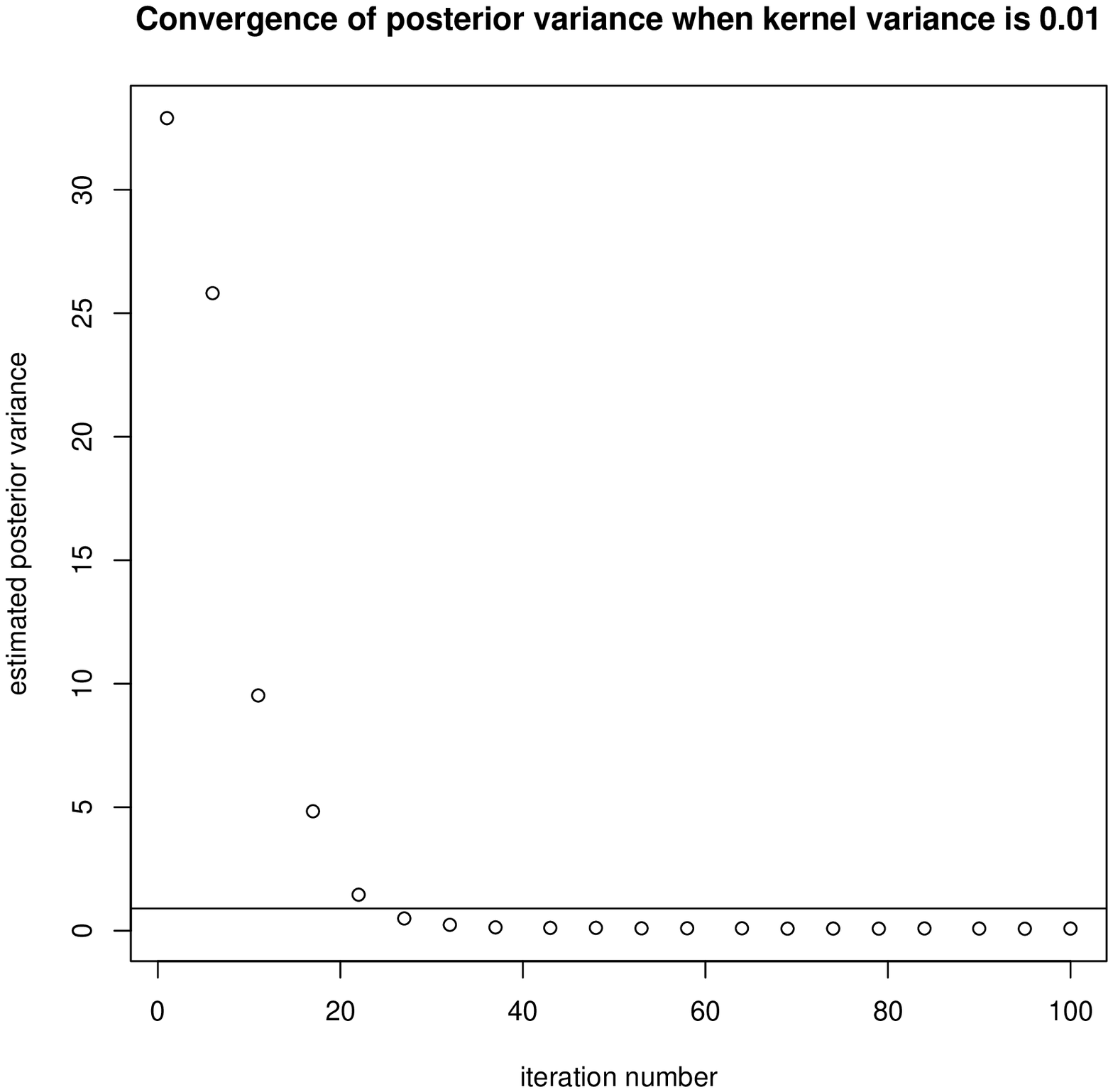}
\caption[]{Application of the ABC-PRC algorithm
 with kernel variance of 0.01. The horizontal line shows
 the theoretical value.}\label{fig:smaller-kern-converge}
\end{figure}

Using an even narrower kernel of 0.01, we can see even greater
discrepancies, as illustrated in Figures \ref{fig:smaller-kern} and
\ref{fig:smaller-kern-converge}. In this case the posterior variance
to which the method converges is around 0.094, almost one-tenth of
the correct value.

\begin{figure}[tb]
\includegraphics[width=6in]{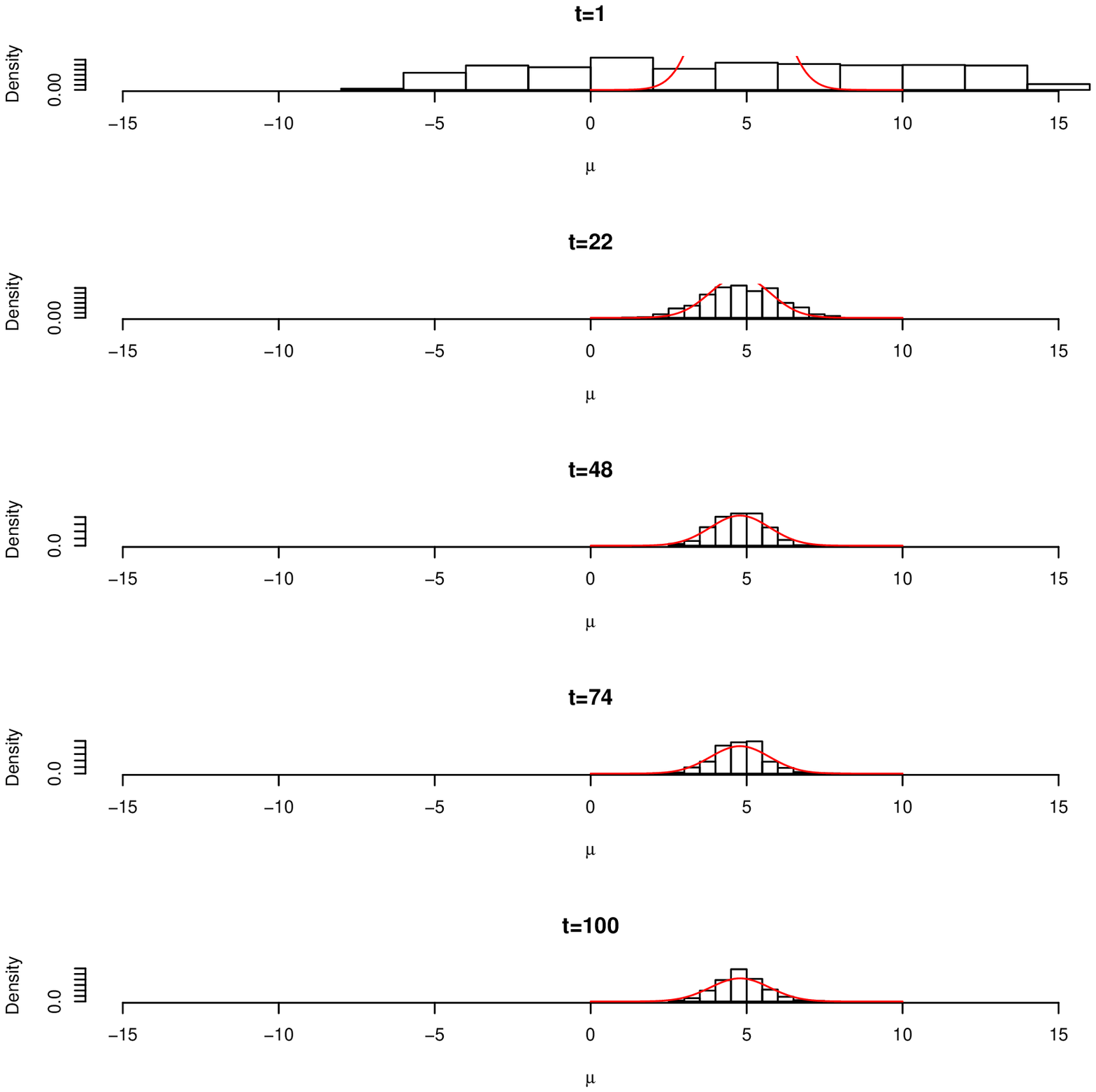}
\caption[]{Application of the ABC-PRC algorithm
 with kernel variance of 1}\label{fig:medium-kern}
\end{figure}

\begin{figure}[tb]
\centering
\includegraphics[width=4in]{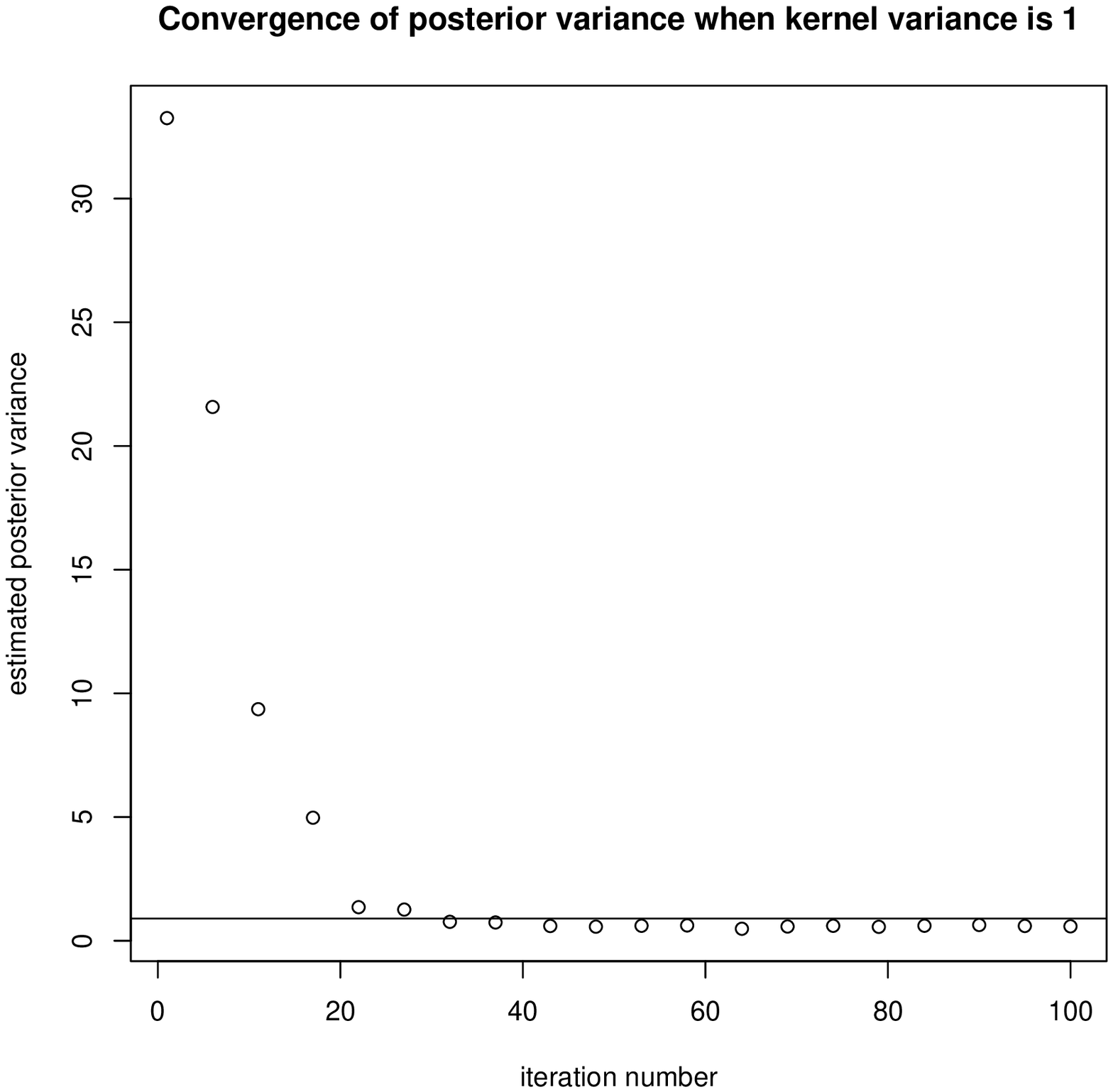}
\caption[]{Application of the ABC-PRC algorithm
 with kernel variance of 1. The horizontal line shows
 the theoretical value.}\label{fig:medium-kern-converge}
\end{figure}

Increasing the variance of the kernel to 1, we see that the
estimated posterior variance, now around 0.59, becomes closer to the
true value of 0.9 (Figure \ref{fig:medium-kern-converge}), and,
superficially, the posterior distribution looks similar to the
theoretical distribution (Figure \ref{fig:medium-kern}.

\begin{figure}[tb]
\includegraphics[width=6in]{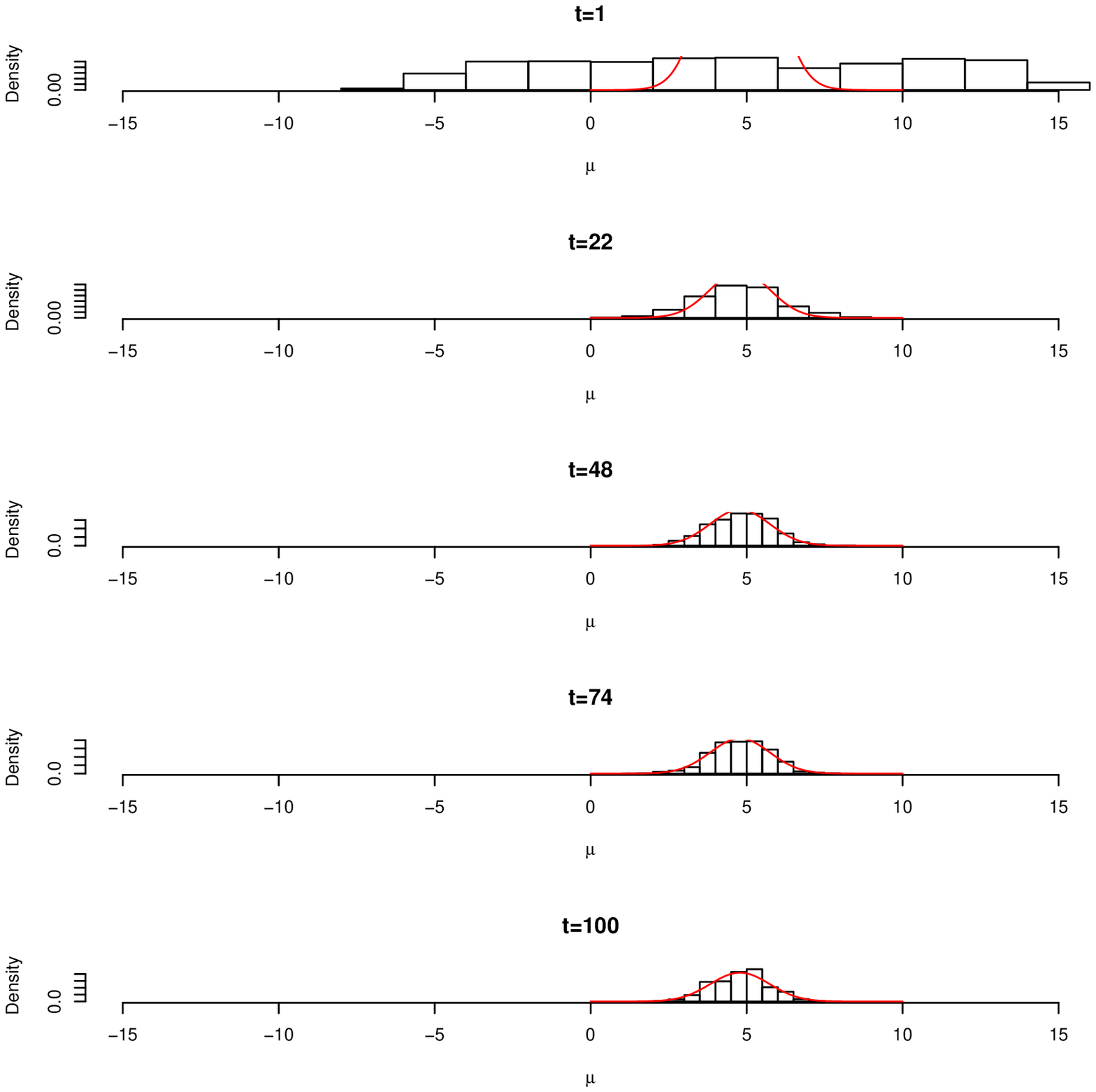}
\caption[]{Application of the ABC-PRC algorithm
 with kernel variance of 10}\label{fig:wide-kern}
\end{figure}

\begin{figure}[tb]
\centering
\includegraphics[width=4in]{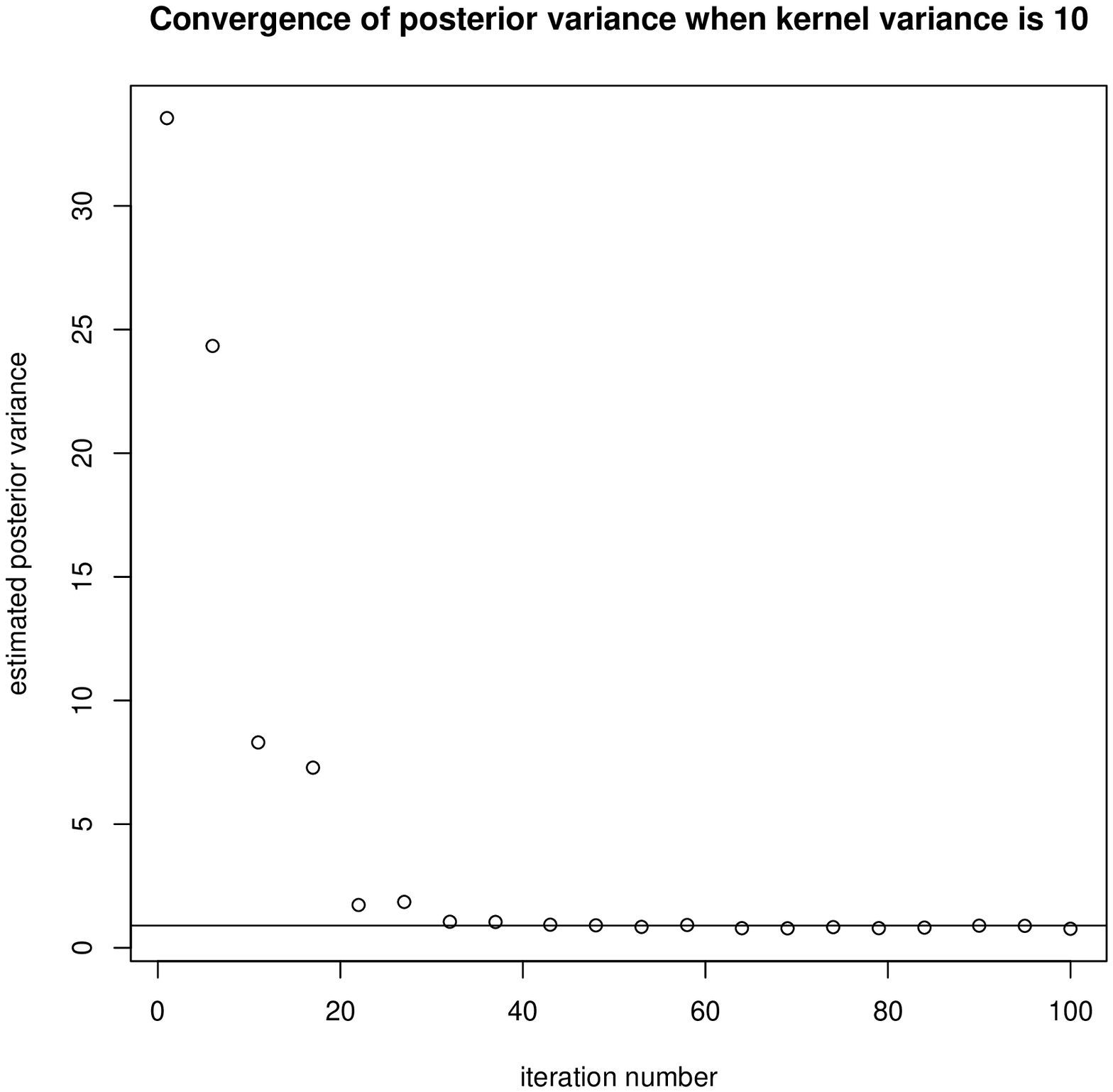}
\caption[]{Application of the ABC-PRC algorithm
 with kernel variance of 10. The horizontal line shows
 the theoretical value.}\label{fig:wide-kern-converge}
\end{figure}

Finally with the variance of the kernel at 10, the estimated
posterior variance is around 0.84 (Figure
\ref{fig:wide-kern-converge}), still lower than the theoretical
value, and the posterior distribution appears very similar to the
theoretical distribution (Figure \ref{fig:wide-kern}).

\section{Solution}

These results suggest that a good way to look at the  ABC-PRC
algorithm of Sisson \emph{et al.} (2007) is as a successive series
of applications of the rejection algorithm to random variables drawn
from a prior distribution that is a convolution of the smoothing
kerneal and the realized variables from the previous rejection
round. If this kernel is large relative to the posterior
distribution the method will \emph{appear} to work because it is
similar to drawing the variables from a uniform prior. However it
can be seen that with a lower kernel variance, the posterior
distribution is badly estimated. Viewed in this way, it is clear
that \emph{the weights do matter}, and the appropriate correction,
at least for the Gaussian kernel used here, is to compute the
weights, $W_{(t),i}$ for the $i$th particle, from the reciprocal of
the (unnormalised) kernel density estimate as
$$
W_{(t),i} = \frac{1}{\sum_{j=0}^N p(\theta_{(t),i} |
\theta_{(t-1),j},\xi^2)}
$$
where $ p(\theta_{(t),i} | \theta_{(t-1),j},\xi^2)$ is a Gaussian
with mean $\theta_{(t-1),j}$ and variance of the kernel that is
used, $\xi^2$.

\begin{figure}[tb]
\includegraphics[width=6in]{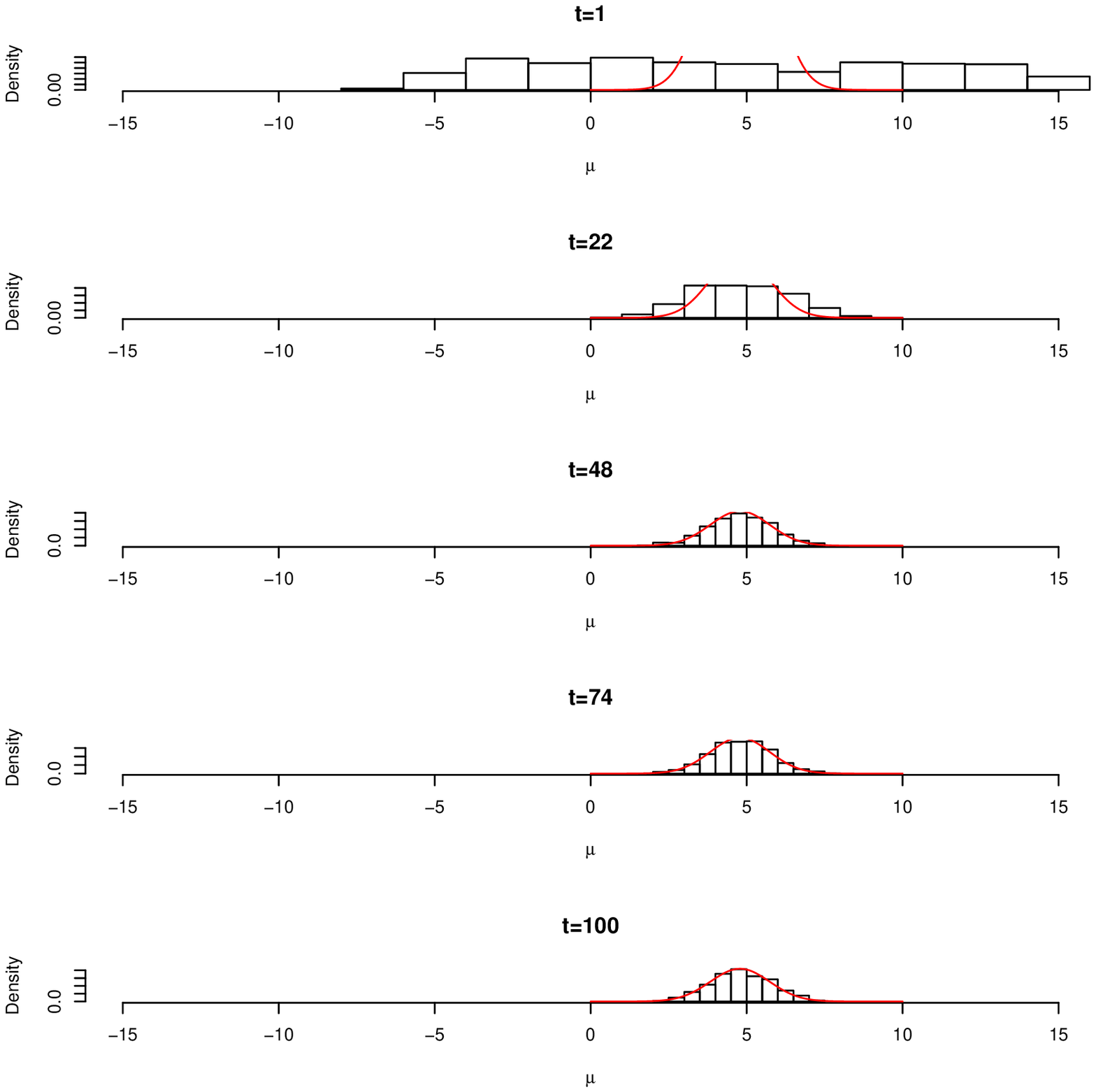}
\caption[]{Application of the corrected ABC-PRC algorithm
 with kernel variance of 0.01. Compare with the uncorrected version
 in Figure \ref{fig:smaller-kern}}\label{fig:correct-kern}
\end{figure}

\begin{figure}[tb]
\centering
\includegraphics[width=4in]{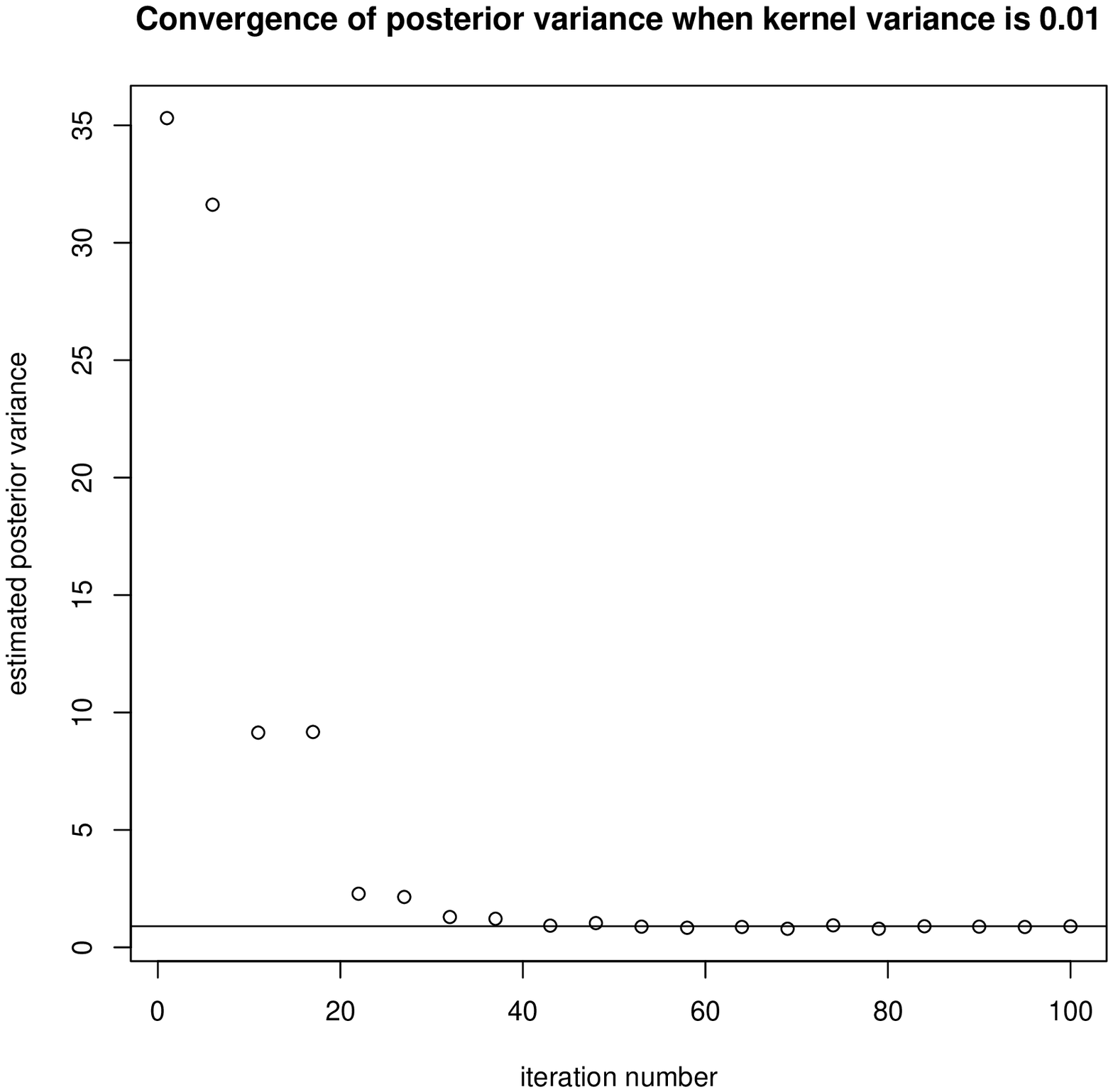}
\caption[]{Application of the corrected ABC-PRC algorithm
 with kernel variance of 0.01. The horizontal line shows
 the theoretical value. Compare with the uncorrected version
 in Figure \ref{fig:smaller-kern-converge}}\label{fig:correct-kern-converge}
\end{figure}

Figures \ref{fig:correct-kern} and \ref{fig:correct-kern-converge}
show the results of applying this corrected version of the ABC-PRC
algorithm to the example with a kernel of variance 0.01, illustrated
in Figures \ref{fig:smaller-kern} and
\ref{fig:smaller-kern-converge}. In this case the estimate variance
converges to around 0.88, which compares favourably with the
uncorrected ABC-PRC algorithm even when a large kernel is used.

In conclusion, it would appear that the ABC-PRC algorithm of Sissons
\emph{et al.} is wrong and should not be used. It can, however, be
corrected by the addition of a computationally trivial weighting
scheme.

\clearpage

\section{Appendix}

\textbf{The R Function for performing the original ABC-PRC}

{\tiny
\begin{verbatim}
abc.smc1a <- function(npart,niter,unif.lo,unif.hi,y,sigma2,eps,kern.mean,kern.var)
{
# npart is the number of particles
# niter is the number of smc iterations
# unif.lo is the lower limit of the uniform distribution
# unif.hi is the upper limit of the uniform distribution
# y is the real data - we compute the only summary stat - the sample mean - from this
# sigma2 is the known variance.
# eps is the vector of tolerances - the euclidean distance must be less than this

    if(length(eps) != niter)stop("eps is wrong length")
    if(niter < 20)nscore <- niter
    else nscore <- 20
    p.history <- matrix(nrow=nscore,ncol=npart)
    pt <- round(seq(1,niter,length=nscore))
    nsamp <- length(y)
    ymean <- mean(y)
    v1 <- runif(npart,unif.lo,unif.hi)
    k <- 1
    for(j in 1:niter){
        vx <- numeric(0)
        while(T){
            vv <- sample(v1,100*npart,replace=T)
            ssvec <- lapply(lapply(vv,rnorm,n=nsamp,sd=sqrt(sigma2)),mean)
            ind <- sqrt((as.numeric(ssvec) - ymean)^2) <= eps[j]
            if(sum(ind) == 0)next;
            vx <- c(vx,vv[ind])
            if(length(vx) >= npart){
                vx <- vx[1:npart]
                break;
            }
        }
        if(pt[k] == j){
            p.history[k,] <- vx;
            k <- k+1
        }
        v1 <- vx + rnorm(npart,kern.mean,sqrt(kern.var))
    }
    p.history
}
\end{verbatim}
}

\noindent \textbf{The R function for performing the corrected ABC-PRC}
{ \tiny
\begin{verbatim}
abc.smc1a.correct <- function(npart,niter,unif.lo,unif.hi,y,sigma2,eps,kern.mean,kern.var)
{
# npart is the number of particles
# niter is the number of smc iterations
# unif.lo is the lower limit of the uniform distribution
# unif.hi is the upper limit of the uniform distribution
# y is the real data - we compute the only summary stat - the sample mean - from this
# sigma2 is the known variance.
# eps is the vector of tolerance - the euclidean distance must be less than this

    if(length(eps) != niter)stop("eps is wrong length")
    if(niter < 20)nscore <- niter
    else nscore <- 20
    p.history <- matrix(nrow=nscore,ncol=npart)
    pt <- round(seq(1,niter,length=nscore))
    nsamp <- length(y)
    ymean <- mean(y)
    v1 <- runif(npart,unif.lo,unif.hi)
    wtvec <- rep(1,npart)
    k <- 1
    for(j in 1:niter){
        vx <- numeric(0)
        while(T){
            vv <- sample(v1,100*npart,replace=T,prob=wtvec)
            ssvec <- lapply(lapply(vv,rnorm,n=nsamp,sd=sqrt(sigma2)),mean)
            ind <- sqrt((as.numeric(ssvec) - ymean)^2) <= eps[j]
            if(sum(ind) == 0)next;
            vx <- c(vx,vv[ind])
            if(length(vx) >= npart){
                vx <- vx[1:npart]
                break;
            }
        }
        if(pt[k] == j){
            p.history[k,] <- vx;
            k <- k+1
        }
        v1 <- vx + rnorm(npart,kern.mean,sqrt(kern.var))
       for(jj in 1: npart)wtvec[jj] <- 1.0/sum(dnorm(v1[jj],vx,sqrt(kern.var)))
    }
    p.history
}
\end{verbatim}
}
\end{document}